\documentclass[12pt,english,floatfix,superscriptaddress,aps,prd,preprint]{revtex4}
\usepackage[utf8]{inputenc}
\usepackage{amsmath}
\usepackage{amssymb}
\usepackage{amsbsy}
\usepackage{amsfonts}
\usepackage{amsopn}
\usepackage{amstext}
\usepackage{graphicx}
\usepackage{amssymb}
\usepackage{amsfonts}
\usepackage{amsmath}
\usepackage{amsmath,amsthm,amsfonts,amssymb}
\usepackage[mathcal]{eucal}
\usepackage{mathrsfs}
\usepackage{graphicx}
\usepackage[english]{babel}
\usepackage{color}
\usepackage{esint}
\usepackage[dvips]{epsfig}
\usepackage[dvips]{graphicx}
\usepackage{float}
\usepackage{units}
\usepackage{textcomp}

\usepackage{hyperref}
\usepackage{slashed}

\newcommand{\ie}{\begin{equation}}
\newcommand{\fe}{\end{equation}}
\newcommand{\se}{\begin{eqnarray}}
\newcommand{\ff}{\end{eqnarray}}

\usepackage{xcolor}
\usepackage{soul}

\begin{document}

\title{Ringlike vortices in a logarithmic generalized Maxwell theory}


\author{F. C. E. Lima}
\email{cleiton.estevao@fisica.ufc.br}
\affiliation{Universidade Federal do Cear\'a (UFC), Departamento de F\'isica,\\ Campus do Pici, Fortaleza - CE, C.P. 6030, 60455-760 - Brazil.}


\author{C. A. S. Almeida}
\email{carlos@fisica.ufc.br}
\affiliation{Universidade Federal do Cear\'a (UFC), Departamento de F\'isica,\\ Campus do Pici, Fortaleza - CE, C.P. 6030, 60455-760 - Brazil.}

\date{\today}

\begin{abstract}
We investigate the presence of vortex structures in a Maxwell model with a logarithmic generalization. This generalization becomes important because it generates stationary field solutions in models that describe the dynamics of a scalar field. In this work, we will choose to investigate the dynamics of the complex scalar field with the gauge field governed by Maxwell term. For this, we will investigate the Bogomol'nyi equations to describe the static field configurations. Then, we show numerically that the complex scalar field solutions that generate minimum energy configurations have internal structures. Finally, assuming a planar vision, the magnetic field and the density energy show the interesting feature of the ringlike vortex.
\end{abstract}

\maketitle


\section{Introduction}
Since 1976, when Bogomol'nyi in his seminal paper, discussed an innovative way for the stability of classical solutions \cite{Bogomolnyi}, the study of topological defects have attracted the attention of several researchers \cite{Ghosh, Lima, Sales, Cunha, Dionizio, Dionizio1, Matheus}. Indeed,  topological structures have aroused a broad interest because these solutions can be used to represent particles and cosmic objects, such as cosmic strings \cite{Nielsen}.

Several approaches to the study of defects have been applied in recent years to various systems. The large amount of works dedicated to the theme is due to the fact that each model presents a different dynamic depending on how the fields can be coupled. In this way, the study of topological defects that describe particles and cosmic objects helps everyone to understand a little more about the universe around us. Let us mention only very recent works about these structures. It was proposed by Lima {\it et al.} an interesting way how to describe solutions of kink defects that can be continuously transformed into compactons solutions \cite{Lima, Lima1}. Others have turned their attention to studies properties of finite temperature defects \cite{Dionizio2}. In the context of braneworlds, solutions with topological defects would also appear \cite{Almeida}.

As well noted in ref. \cite{Dionizio3}, topological objects that appear in nonlinear dynamics systems known as kink, vortex, and monopoles, are related to the spatial dimensions of the model. For example, in $(1+1)D$, we often get well-known kink solutions. Meanwhile, in $(2+1)D$, we have topological structures called vortex, and in the case of $(3+1)D$ of spacetime, we can obtain the solutions known as monopoles. It is worth noting that the simplest of these possible structures are kink defects that can even be obtained in models such as $\lambda\phi^{4}$ with the spontaneous breaking of symmetry. This attractive matter can be addressed in the excellent reviews of Rajaraman \cite{Rajaraman}, Vachaspati \cite{Vachaspati}, and Manton \cite{Manton}.

Study of planar structures like vortex has been investigated since the beginning of the theory proposed by Bogomol'nyi-Prasad-Sommerfield \cite{Bogomolnyi, Prasad, Casana, Casana1}. However, in recent years, these planar structures have been investigated in generalized models such as \cite{Losano, Dionizio4}. These generalizations are quite interesting because they can generate different characteristics \cite{Lee, Babichev}. A new point that arouses the attention of many adepts of the subject is to study the dynamics of scalar, vector, and complex fields coupled with Gauge fields in generalized models, as the work mentioned in ref. \cite{Dionizio3}. The fact is that when making these generalizations it is always possible to investigate new properties that describe the dynamic of field, even if in principle, this field is coupled with other fields such as the gauge \cite{Antonio, Jackiw}.

In our work we will introduce for the first time a logarithmic interaction in Abelian topological models.  Tracking back to Rosen \cite{rosen,rosen1}, with their works in relativistic theories and Bialynicki-Birula and Mycielski \cite{birula} in the context of Schroedinger theories, logarithmic potentials were readdressed recently in the context of (1+1) relativistic models by  Belendryasova {\it et. al.} \cite{E}. In this latter work, the authors obtained kinks and other soliton-like solutions called \textit{gaussons} for their Gaussian shape.

On the other hand, we can find some applications of the logarithmic potential in (3+1) models. Indeed, when we consider electric charge as a self-consistent configuration of an electromagnetic field interacting with a physical vacuum, the nonlinearity could be extremely important since the system is effectively described as a logarithmic quantum Bose liquid \cite{Vladimir}. Also, models with the so called gausson logarithmic term were used to describe theories of quantum gravity \cite{Konstantin2} as well as quantum effects in nonlinear quantum theory \cite{Konstantin1}.

In this work,  our goal is investigate the influence of a new coupling between scalar and gauge field in the topological solutions of the model. This new coupling comes from a generalization in the Maxwell term. So, we define the model and investigated their stationary solutions using the BPS approach. Then, we analyze the corresponding planar shape for the magnetic field and Bogomol'nyi energy density of the model. Subsequently, we study the particular case of an unmodified Maxwell term and compare the results with the generalized case. Finally, we present some final considerations about the model studied.

\section{The logarithmic generalized Maxwell model} 

We started our investigation considering the generalized Lagrangian density in flat space-time in $(2+1)D$ given by
\begin{equation}
\label{general}
\mathcal{L}=-\frac{1}{4}G(|\phi|)F_{\mu\nu}F^{\mu\nu}+|D_{\mu}\phi|^{2}-V(|\phi|).
\end{equation}

This generalized Lagrangian was introduced for the first time by Lee and Nam \cite{lee1} in order to find solitons solutions of an Abelian Higgs model. Soon after, Bazeia \cite{bazeia1}, used it to find vortices for different choices of the function $ G(|\phi|)$.  This general function is sometimes called a dielectric permeability \cite{lee1,bazeia1,ghosh1}. In principle, this function, which is a function of the complex scalar field module, should be a non-negative function of the scalar field module.
Inspired by the work of Belendryasova {\it et. al.} \cite{E}, we choose the function $ G(|\phi|)$ as a logarithmic function of $\phi$.

As mentioned above, a logarithmic interaction was already used by several authors. But all of them, used a term describing a nonlinear self-interaction of scalar fields. For instance, the most recent work consider a relativistic scalar theory in (1+1) dimensions for the study of topological structures generating kink and antikink solutions \cite{E}. Here, we propose the function $ G(|\phi|)$ as a logarithmic  one, which leads to a non-linear coupling of scalar fields to a gauge field. The Lagrangian proposed is

\begin{equation}
\label{log}
\mathcal{L}=\frac{|\phi|^{2}}{4\lambda^{2}}\bigg[\text{ln}\bigg(\frac{|\phi|^{2}}{\vartheta^{2}}\bigg)-1\bigg]F_{\mu\nu}F^{\mu\nu}+|D_{\mu}\phi|^{2}-V(|\phi|),
\end{equation}
where $\lambda$ and $\vartheta$ are parameters that adjusts the canonical dimension of the model. Here the scalar field $\phi$ is complex and $|\phi|^{2}=\phi\bar{\phi}$. The notation $\bar{\phi}$ representing the conjugate complex of $\phi$. $A_{\mu}$ is the Abelian gauge field, $F_{\mu\nu}=\partial_{\mu}A_{\nu}-\partial_{\nu}A_{\mu}$ is the electromagnetic tensor. The potential is given by $V(|\phi|)$ and the metric signature is $\eta_ {\mu\nu}=$diag$(+, -, -)$. We also consider $\hbar=c=1$.

The covariant derivative is defined as
\begin{equation}
D_{\mu}\phi=\partial_{\mu}\phi+ieA_{\mu}\phi.
\end{equation}

Investigating the equations of motion associated with the model, we obtain:
\begin{equation}
\label{motion}
D_{\mu}D^{\mu}\phi+\frac{\phi}{2|\phi|}\bigg[-\frac{|\phi|}{2\lambda^{2}}\text{ln}\bigg(\frac{|\phi|^{2}}{\vartheta^{2}}\bigg)F_{\mu\nu}F^{\mu\nu}+V_{|\phi|}\bigg]=0
\end{equation}
and
\begin{equation}
\label{current}
\partial_{\mu}\bigg\{-\frac{|\phi|^{2}}{\lambda^{2}}\bigg[\text{ln}\bigg(\frac{|\phi|^{2}}{\vartheta^{2}}\bigg)-1\bigg]F^{\mu\nu}\bigg\}=J^{\nu}.
\end{equation}

We define the current as
\begin{equation}
J_{\mu}=ie(\bar{\phi}D_{\mu}\phi-\phi\overline{D_{\mu}\phi}),
\end{equation}
and $V_{|\phi|}=\partial V/\partial |\phi |$.

Considering stationary field configurations and defining the component $\nu=0$ in the Eq. (\ref{current}), we observe that Gauss's law is satisfied for the component of the Abelian field $A_{0}=0$. This implies the possible existence of electrically neutral vortex.

Through the translational symmetry of space-time in the model, we build the energy-momentum tensor as
\begin{equation}
\label{Tmn}
T_{\mu\nu}=-\frac{|\phi|^{2}}{\lambda^{2}}\bigg[\text{ln}\bigg(\frac{|\phi|^{2}}{\vartheta^{2}}\bigg)-1\bigg]F_{\mu\rho}F^{\rho}\hspace{0.01cm}_{\nu}+\overline{D_{\mu}\phi}D_{\nu}\phi+\overline{D_{\nu}\phi}D_{\mu}\phi-\eta_{\mu\nu}\mathcal{L}.
\end{equation}

In order to investigate the vortex solutions of the model we will use the ansatz
\begin{equation}
\label{ansatz}
\phi=g(r)\text{e}^{in\theta} \hspace{0.5cm} \text{e} \hspace{0.5cm} \vec{A}=-\frac{1}{er}[a(r)-n]\hat{\theta},
\end{equation}
with $n \, \in \, \mathbb{Z}$ and representing the vorticity of the system.

Considering the ansatz (\ref{ansatz}), we get the expression that describes the magnetic field, that is,
\begin{equation}
\label{magnetic}
\vec{B}=\vec{\nabla}\times\vec{A}\rightarrow B\equiv ||\vec{B}||=-\frac{a'(r)}{er}=-F_{12},
\end{equation}
where the prime in $a'(r)$ denotes the derivative with respect to the radial coordinate $r$. With this in mind, we note that the magnetic field flux $\Phi_{B}$ is given by
\begin{equation}
\label{flux}
\nonumber
\Phi_{B}=\int_{_S} \int \, \vec{B} \cdot d\vec{S}=-\int_{_S}\int \, F^{12} \hat{z}\cdot r\, dr\, d\theta \hat{z}=-\int_{0}^{2\pi}\int_{0}^{\infty} F^{12} r\, dr\, d\theta, 
\end{equation}
or
\begin{equation}
\Phi_{B}=-\frac{2\pi}{e}\int_{0}^{\infty}\frac{da(r)}{dr}dr=-\frac{2\pi}{e}[a(\infty)-a(0)].
\end{equation}

From now on, we assume that the asymptotic behaviour for the ansatz (\ref{ansatz}) is
\begin{align}
\label{condition} \nonumber
& g(0)=0, \hspace{1cm} a(0)=n, \\
& g({\infty})=\nu, \hspace{1cm} a(\infty)=0.
\end{align}

Substituting (\ref{condition}) in (\ref{flux}), we arrive at the expression
\begin{equation}
\Phi_{B}=\frac{2\pi n}{e}.
\end{equation}
Therefore, the magnetic flux of the model is quantized.

Considering the ansatz (\ref{ansatz}), the equations of motion are now rewritten as
\begin{equation}
(r g'(r))'=\frac{a(r)^{2}g(r)}{r}-\frac{g(r)a'(r)}{2\lambda^{2}e^{2}r}\text{ln}\bigg(\frac{g(r)^{2}}{\vartheta^{2}}\bigg)+\frac{1}{2}r V_{g(r)},
\end{equation}
and
\begin{equation}
\bigg\{\frac{g(r)^{2}a'(r)}{\lambda^{2}er}\bigg[\text{ln}\bigg(\frac{g(r)^{2}}{\vartheta^{2}}\bigg)-1\bigg]\bigg\}^{'}=-\frac{2ea(r)g(r)^{2}}{r}.
\end{equation}

Now, we will turn our attention to investigating the energy density of the model. Considering that a component $T_{00}$ of the energy-momentum tensor (\ref{Tmn}) represents the energy density of the model, we have 

\begin{align}
\label{energy}
\mathcal{E}= -\frac{g(r)^{2}a'(r)^{2}}{2e^{2}r^{2}\lambda^{2}}\bigg[\text{ln}\bigg(\frac{g(r)^{2}}{\vartheta^{2}}\bigg)-1\bigg]+g'(r)^{2}+\frac{a(r)^{2}g(r)^{2}}{r^{2}}+V(g).
\end{align}

Seeking to satisfy the BPS limit, we rearranged the previous expression. Than, we can write,
\begin{align}\label{energy1} \nonumber
\mathcal{E}=&-\frac{g(r)^{2}}{2\lambda^{2}}\bigg[\text{ln}\bigg(\frac{g(r)^{2}}{\vartheta^{2}}\bigg)-1\bigg]\bigg\{\frac{a'(r)}{er}\mp\frac{e\lambda^{2}(\nu^{2}-g(r)^{2})}{g(r)^{2}\bigg[\text{ln}\bigg(\frac{g(r)^{2}}{\vartheta^{2}}\bigg)-1\bigg]}\bigg\}^{2}+\bigg(g'(r)\mp\frac{a(r)g(r)}{r}\bigg)^{2} \\ & +V+\frac{e^{2}\lambda^{2}}{2g(r)^{2}}\frac{(\nu^{2}-g(r)^{2})^{2}}{\bigg[\text{ln}\bigg(\frac{g(r)^{2}}{\vartheta^{2}}\bigg)-1\bigg]}\mp\frac{1}{r}[a(r)(\nu^{2}-g(r)^2)]'.
\end{align}

If the choice of potential is such that
\begin{equation}
V=-\frac{e^{2}\lambda^{2}}{2g(r)^{2}}\frac{(\nu^{2}-g(r)^{2})^{2}}{\bigg[\text{ln}\bigg(\frac{g(r)^{2}}{\vartheta^{2}}\bigg)-1\bigg]},
\end{equation}
the previous expression is reduced significantly.

Considering that the energy of the model is the integration of energy density in all space, the energy of the model takes the form,
\begin{align}
\label{energy2}\nonumber
E=&-2\pi\int_{0}^{\infty}\, r dr \, \frac{g(r)^{2}}{2\lambda^{2}}\bigg[\text{ln}\bigg(\frac{g(r)^{2}}{\vartheta^{2}}\bigg)-1\bigg]\bigg\{\frac{a'(r)}{er}\mp\frac{e\lambda^{2}(\nu^{2}-g(r)^{2})}{g(r)^{2}\bigg[\text{ln}\bigg(\frac{g(r)^{2}}{\vartheta^{2}}\bigg)-1\bigg]}\bigg\}^{2} \\ &+2\pi\int_{0}^{\infty}\, r dr \, \bigg(g'(r)\mp\frac{a(r)g(r)}{r}\bigg)^{2}+E_{BPS},
\end{align}
where we have,
\begin{align}
\label{e}
E_{BPS}=\mp 2\pi\int_{0}^{\infty}\, dr \, [a(r)(\nu^{2}-g(r)^{2})]'=2\pi|n|\nu^{2}.
\end{align}

In other words, we notice that the energy is limited by $E_{BPS}$, i. e., 
\begin{equation}
\label{Bogomol'nyi}
    E\geq E_{BPS}.
\end{equation}

Therefore, if the solution obeys the first order equations
\begin{align}
\label{b}
    g'(r)=\pm \frac{a(r)g(r)}{r};
\end{align}

\begin{align}
\label{b1}
\frac{a'(r)}{er}=\pm\frac{\lambda^{2}e}{g(r)^{2}}\frac{(\nu^{2}-g(r)^{2})}{\bigg[\text{ln}\bigg(\frac{g(r)^{2}}{\vartheta^{2}}\bigg)-1\bigg]},
\end{align}
then, the BPS limit is saturated and the inequality (\ref{Bogomol'nyi}) becomes an equality, that is, $E=E_{BPS}$.

Decoupling the equations (\ref{b}) and (\ref{b1}), we have:
\begin{align}
g''(r)-\frac{g'(r)^{2}}{g(r)}+\frac{g'(r)}{r}-\frac{\lambda^{2}e^{2}(\nu^{2}-g(r)^{2})}{g(r)\bigg[\text{ln}\bigg(\frac{g(r)^{2}}{\vartheta^{2}}\bigg)-1\bigg]}=0.
\end{align}

In this context, we obtain that the BPS energy density is given by:
\begin{align}
\label{Eb}
\mathcal{E}_{BPS}=-\frac{g(r)^{2}a'(r)^{2}}{e^{2}\lambda^{2}r^{2}}\bigg[\text{ln}\bigg(\frac{g(r)^{2}}{\vartheta^{2}}\bigg)-1\bigg]+2g'(r)^{2}.
\end{align}

\subsection{Numerical results}

\begin{figure}[ht!]
\centering
\includegraphics[scale=0.5]{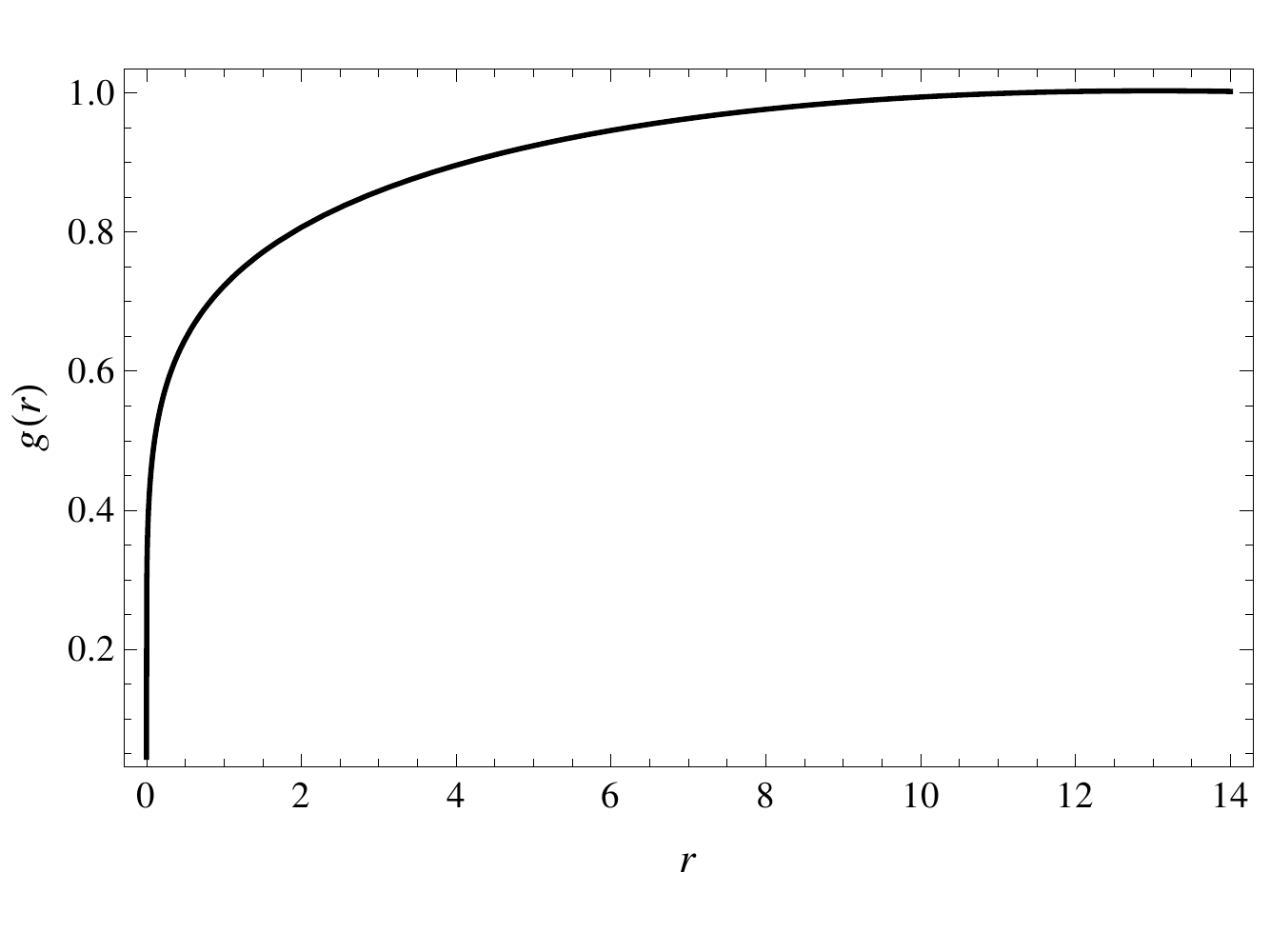}
\includegraphics[scale=0.5]{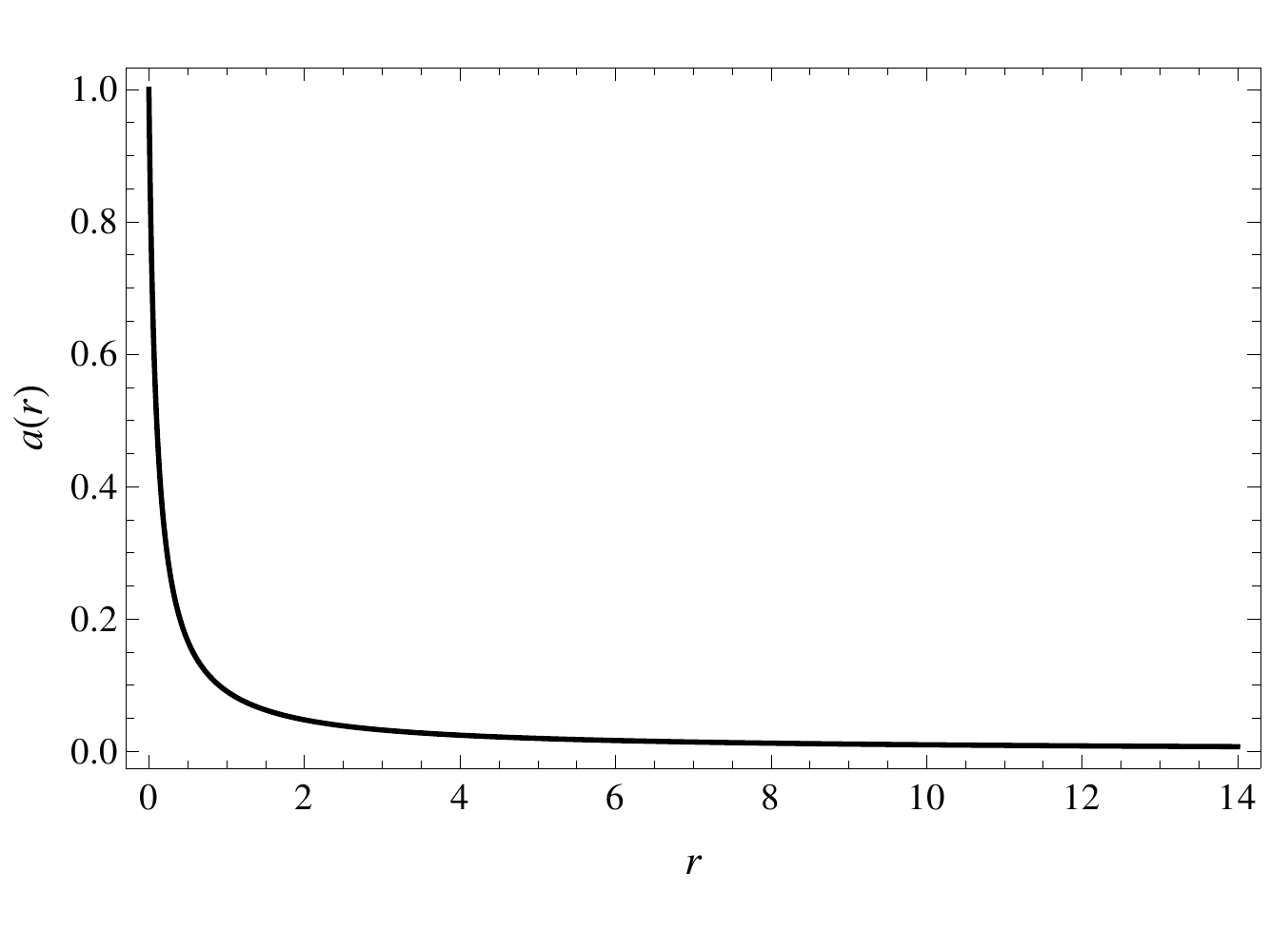}
\vspace{-20pt}
\caption{Numerical solution of the scalar field with vacuum state $\nu=1$, vorticity $n=1$, couplings $\lambda=10^{- 3}$ and $ \vartheta=10^{-5}$ (left side). Behaviour of the gauge field associated with the scalar field (right side).} \label{fig1}
\end{figure}

Now let us turn out attention to the study of the numerical solutions of the vortex structures of our model. Initially, we investigated the topological solutions with the boundary conditions expressed in Eqs. (\ref{condition}), where $\nu$ is the vacuum state value. Without loss of generality, we assume that the model vorticity is equivalent the $n=1$. In this way, we obtain the BPS vortices presented in Fig. (\ref{fig1}).

With the numerical solution of the scalar field, we return to BPS equations and investigate the solution for the gauge field. With this in mind, we seek for the $a(r)$ solution. Thus, we obtain the results shown in Fig. (\ref{fig1}).

\begin{figure}[h]
\centering
\includegraphics[scale=0.4]{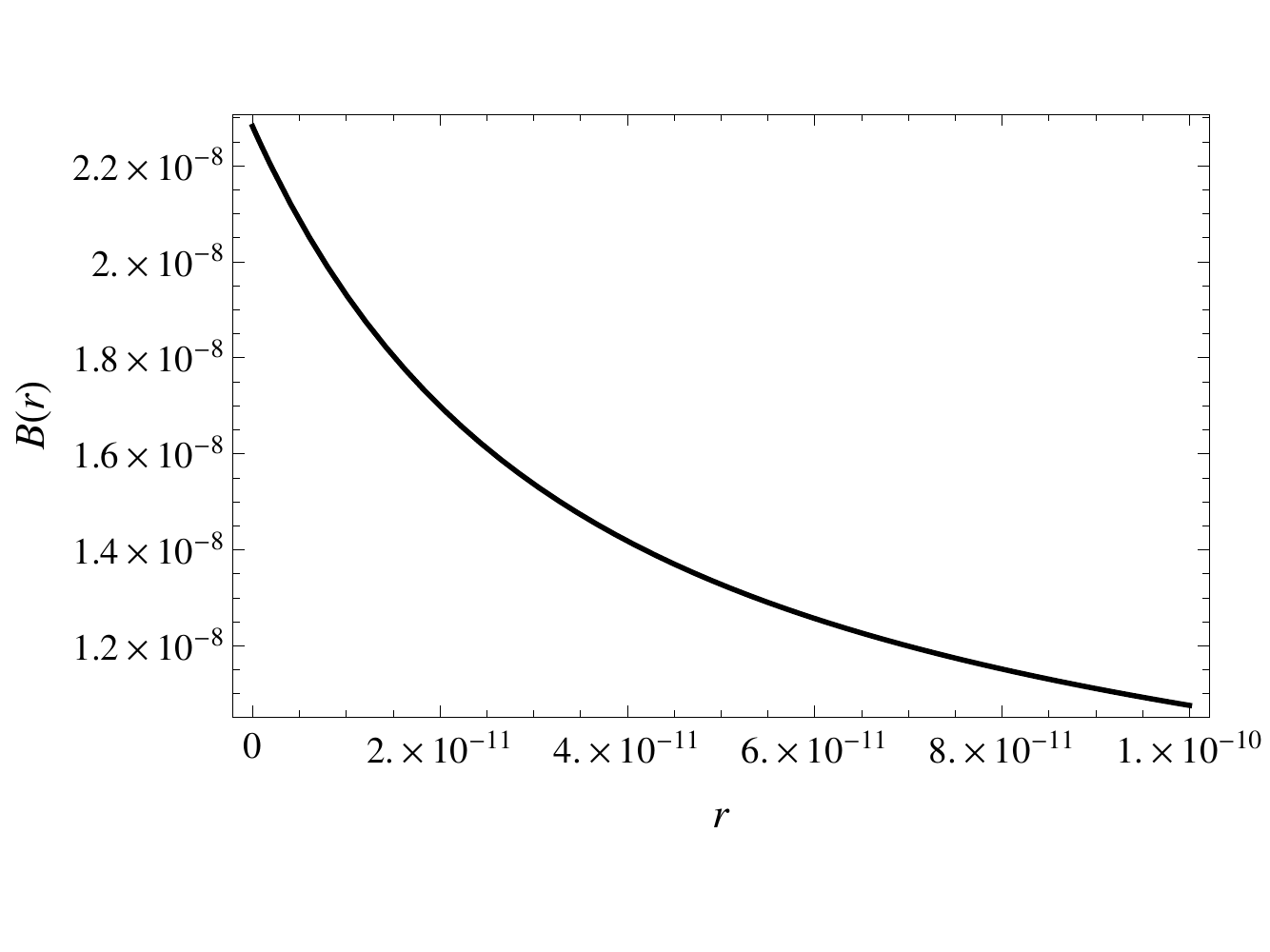}
\includegraphics[scale=0.4]{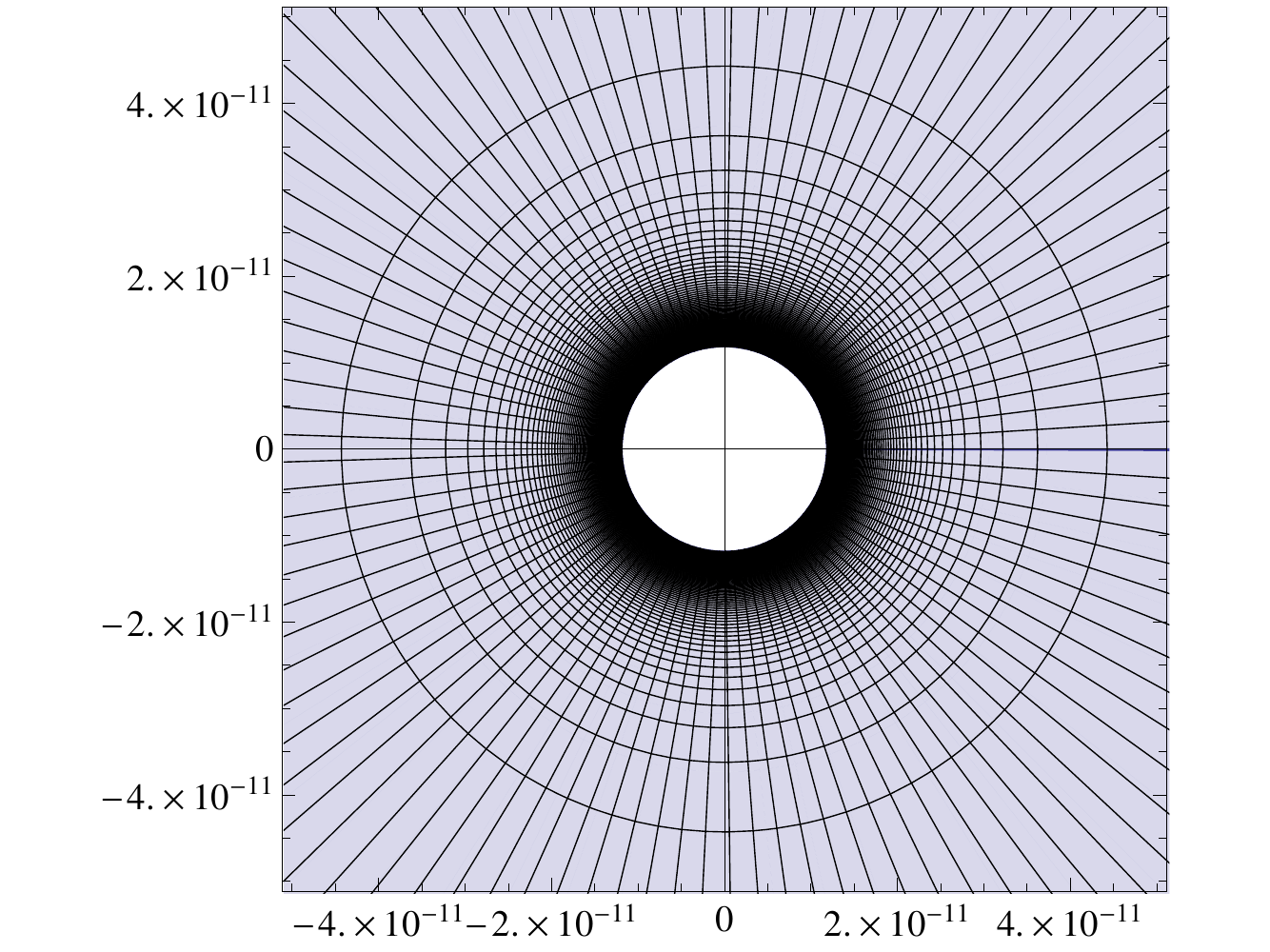}
\caption{The magnetic field of the vortex.} \label{fig3}
\end{figure}
\begin{figure}[h]
\centering
\includegraphics[scale=0.4]{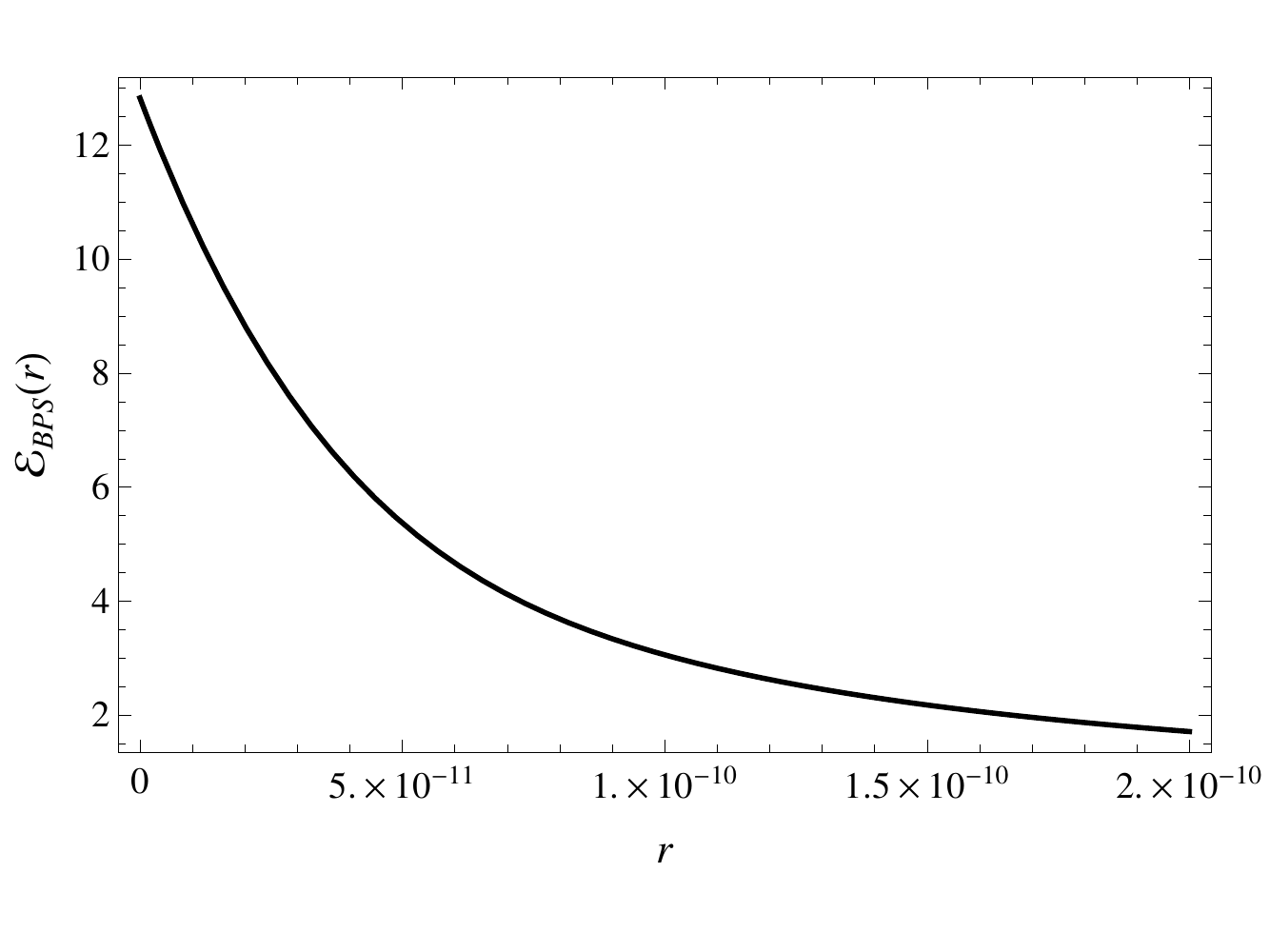}
\includegraphics[scale=0.4]{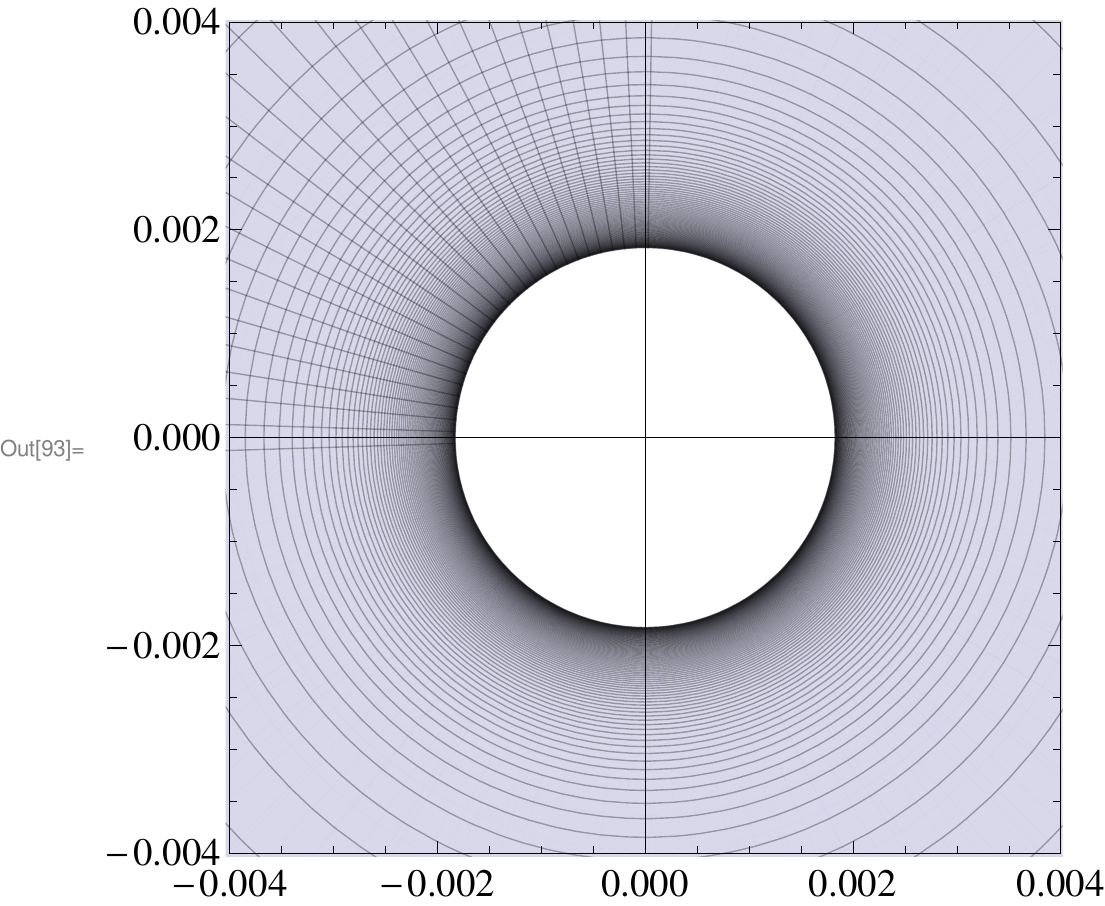}
\caption{The energy density of the vortex.} \label{fig4}
\end{figure}

The magnetic field and energy density of the model can be obtained respectively with the help of the expressions (\ref{magnetic}) and (\ref{Eb}). Therefore, we present the respective graphical results in Figs. (\ref{fig3}) and (\ref{fig4}).

Through the planar representation of the magnetic field in Fig. (\ref{fig3}), we observe the existence of a "big hole" around $r=0$. Also, we realized that the field becomes quite intense in the near of the origin. This can be understood as the contribution of the logarithmic coupling between the scalar field and the gauge field. This interesting feature leads to the so called ringlike vortices \cite{Dionizio3}. This kind of solution has already be found in a Maxwell-Higgs generalized model. Indeed, Bazeia \textit{et. al.}, considered a model with a Lagrangian described by Eq. (\ref{general}). However their function $G(|\phi|)$ is a very particular one \cite{Dionizio3}.


Finally, analyzing the BPS energy density of the model we find the results shown in Fig. (\ref{fig4}). We observed that both the BPS energy density and the magnetic field present similar shapes near the origin and differing only in their magnitude around $r\rightarrow 0$. Furthermore, where the magnetic field is more intense, the BPS energy has smaller magnitude and is more located in space.



\section{The case: generalized Maxwell-Higgs model}

In this case, we consider the generalized Maxwell-Higgs Lagrangian, i. e.,
\begin{equation}
    \mathcal{L}=\frac{\lambda^{2}}{4|\phi|^{2}}\bigg[\text{ln}\bigg(\frac{|\phi|^{2}}{\vartheta^{2}}\bigg)-1\bigg]F_{\mu\nu}F^{\mu\nu}+|D_{\mu}\phi|^{2}-V(|\phi|),
\end{equation}
where $V(|\phi|)$ is a Higgs potential. Than, we have
\begin{equation}
        \mathcal{L}=\frac{\lambda^{2}}{4|\phi|^{2}}\bigg[\text{ln}\bigg(\frac{|\phi|^{2}}{\vartheta^{2}}\bigg)-1\bigg]F_{\mu\nu}F^{\mu\nu}+|D_{\mu}\phi|^{2}-e^{2}(\nu^{2}-|\phi|^{2})^{2}.
\end{equation}

Constructing the motion equations of the model, we obtain the expressions
\begin{equation}
    D_{\mu}D^{\mu}\phi+\frac{\phi}{|\phi|}\bigg(-\frac{\lambda^{2}}{4|\phi|^{2}}\bigg[\text{ln}\bigg(\frac{|\phi|^{2}}{\vartheta^{2}}\bigg)-1\bigg]F_{\mu\nu}F^{\mu\nu}+4e^{2}|\phi||\phi|'(\nu-|\phi|^{2})\bigg)=0,
\end{equation}

\begin{equation}
    -\partial_{\mu}\bigg(\frac{\lambda^{2}}{4|\phi|^{2}}\bigg[\text{ln}\bigg(\frac{|\phi|^{2}}{\vartheta^{2}}\bigg)-1\bigg]F^{\mu\nu}\bigg)=J^{\nu}.
\end{equation}

In order to analyze the structures of BPS vortex, and considering the ansatz of the expression (\ref{ansatz}), we write the equation for the energy as
\begin{equation}
    E=\int\, d^{2}x\, \bigg\{-\frac{g(r)^{2}}{2\lambda^{2}}\bigg[\text{ln}\bigg(\frac{|g(r)^{2}}{\vartheta^{2}}\bigg)-1\bigg]\frac{a'(r)^{2}}{e^{2}r^{2}}+g'(r)+\frac{a(r)^{2}g(r)^{2}}{r^{2}}+e^{2}(\nu^{2}-g(r)^{2})^{2}\bigg\}.
\end{equation}

Reorganizing the energy equation, we obtain that 
\begin{align} \nonumber
    E=&\int\, d^{2}x \bigg\{-\frac{|\phi|^{2}}{4\lambda^{2}}\bigg[\text{ln}\bigg(\frac{|\phi|^{2}}{\vartheta^{2}}\bigg)-1\bigg]\bigg\{\frac{a'(r)}{er}\mp\frac{e\lambda^{2}(\nu^{2}-g(r)^{2})}{g(r)^{2}\bigg[\text{ln}\bigg(\frac{g(r)^{2}}{\vartheta^{2}}\bigg)-1\bigg]}\bigg\}^{2}+\\
    &+\bigg(g'(r)\mp\frac{a(r)g(r)}{r}\bigg)^{2}\mp\frac{1}{r}[a(r)(\nu^{2}-g(r)^{2})]'\bigg\}.
\end{align}

Assuming the BPS limit, we obtain the well-known Bogomol'nyi equations, i. e.,
\begin{equation}
\label{eqb}
    \frac{a'(r)}{er}=\pm\frac{e\lambda^{2}(\nu^{2}-g(r)^{2})}{g(r)^{2}\bigg[\text{ln}\bigg(\frac{g(r)^{2}}{\vartheta^{2}}\bigg)-1\bigg]}
\end{equation}
and
\begin{equation}
\label{eqb1}
 g'(r)=\pm\frac{a(r)g(r)}{r},
\end{equation}
where the energy of the vortex configurations in this case is given by
\begin{equation}
\label{eqb2}
    E=\mp2\pi \int_{0}^{\infty}\, dr\, [a(r)(\nu^{2}-g(r)^{2})]'=2\pi|n|\nu^{2}.
\end{equation}

Note that the expressions (\ref{eqb}), (\ref{eqb1}) and (\ref{eqb2}) are identical to the vortex expressions described in the equations (\ref{e}), (\ref{b}) and (\ref{b1}). Therefore, we emphasize that the case of the generalized Higgs potential will have the same vortices solutions presented in the discussion of the logarithmic generalized Maxwell model presented early. In other words, the numerical results presented in the first section satisfy the case of the generalized Maxwell-Higgs model as well.

\section{The case without generalization}
We turn our attention to the particular case in which the function $ G(|\phi|)=1$. Thus, the Lagrangian density will be reduced to
\begin{equation}
    \mathcal{L}=-\frac{1}{4}F_{\mu\nu}F^{\mu\nu}+|D_{\mu}\phi|^{2}-V(|\phi|).
\end{equation}

In this case, the motion equations are reduced to
\begin{equation}
    D_{\mu}D^{\mu}\phi+\frac{\phi}{2|\phi|}V_{|\phi|}=0
\end{equation}
and
\begin{equation}
    \partial_{\mu}F^{\mu\nu}=J^{\nu}.
\end{equation}

Considering the ansatz (\ref{ansatz}), we rewrite the equations above in terms of the fields $g(r)$ and $a(r)$. Thus, we have 
\begin{equation}
    \frac{1}{r}(rg)'=\frac{a^{2}g}{r}+\frac{a'}{4e^{2}r^{2}}+\frac{1}{2}V_{g},
\end{equation}

\begin{equation}
    r\bigg(\frac{a'}{er}\bigg)'=2eag^{2}.
\end{equation}

With this in mind, we rewrite the expression of the energy density of the model as
\begin{equation}
\label{Eb1}
    \mathcal{E}=\frac{1}{2}\bigg[\frac{a'}{er}\mp e(\nu^{2}-g(r)^{2})\bigg]^{2}+\bigg(g'(r)\mp \frac{ag(r)}{r}\bigg)+V+e^{2}(\nu^{2}-g(r)^{2})^{2}+\mathcal{E}_{BPS}.
\end{equation}
It is worth noting that the expression of the magnetic field remains unchanged in the case without generalization.

Now, taking the potential $V$ as equivalent to the Higgs potential, that is,
\begin{equation}
    V=-e^{2}(\nu^{2}-g(r)^{2})^{2},
\end{equation}
at the BPS limit we will obtain vortex structures with energy described by the equation (\ref{energy2}). In this way, the BPS equations are reduced to
\begin{equation}
\label{higgs}
    g'(r)=\pm\frac{a(r)g(r)}{r}, \hspace{1cm}  \frac{a'(r)}{er}=\pm e(\nu^{2}-g(r)^{2}). 
\end{equation}

\subsection{Numerical results: case without generalization}
\begin{figure}[h!]
\centering
\includegraphics[scale=0.5]{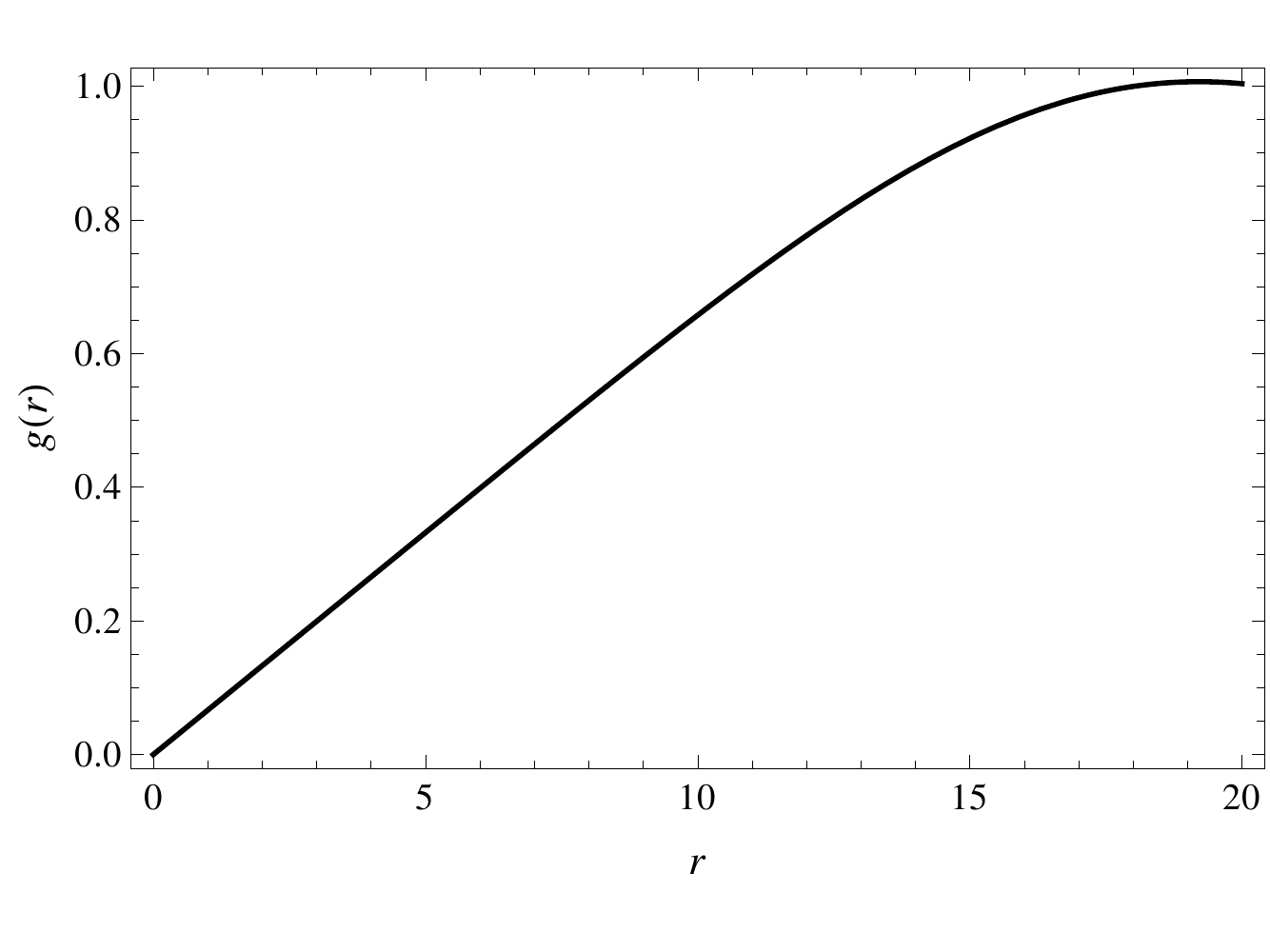}
\includegraphics[scale=0.5]{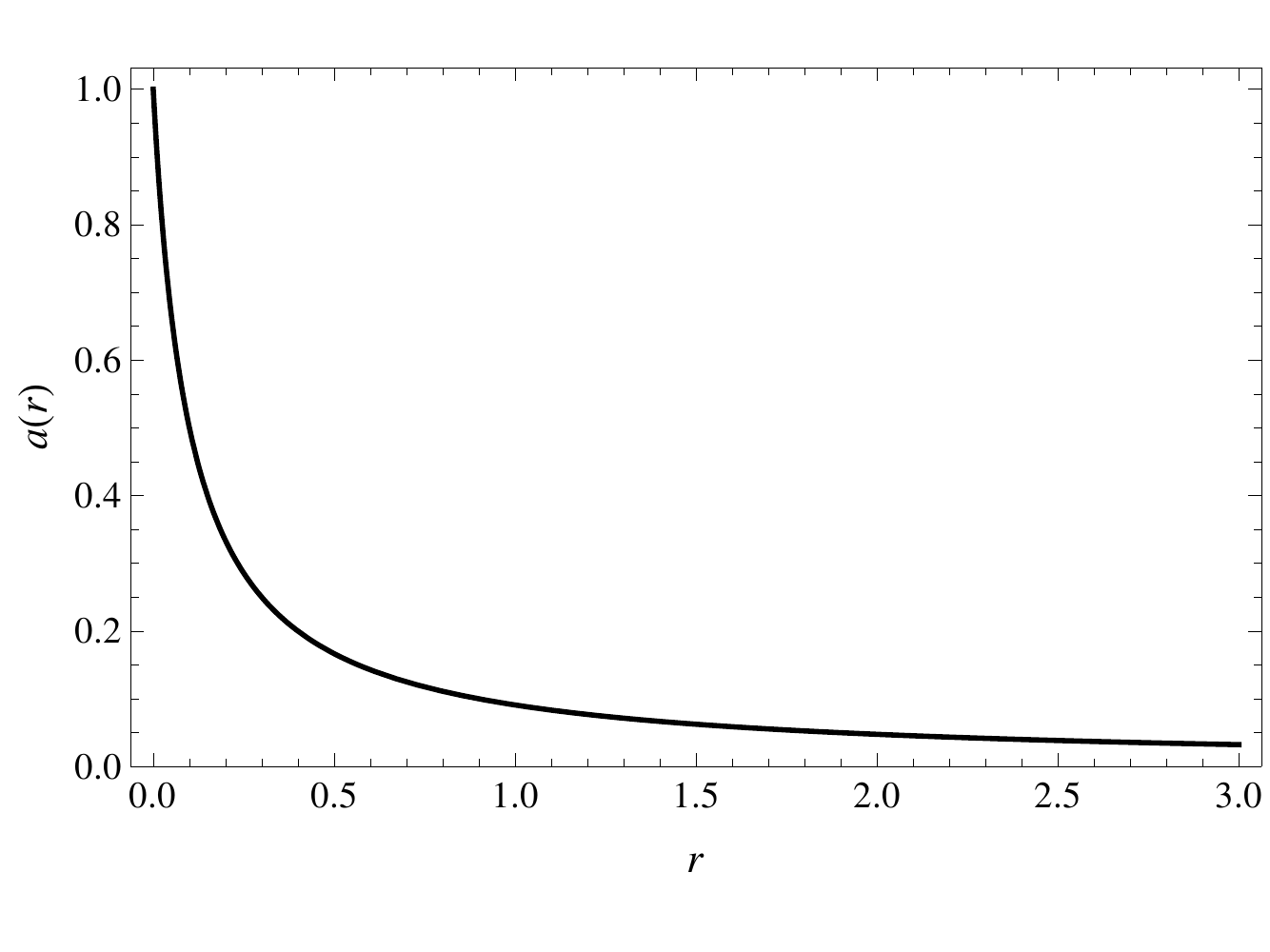}
\vspace{-20pt}
\caption{Numerical solution of the scalar field without generalization (left side). Behaviour of the gauge field associated with the scalar fields when the function $G=1$ (right side).} \label{fig7}
\end{figure}

Using the equation (\ref{higgs}), we will turn our attention to the numerical study of the Higgs field and the gauge field in the model without generalization. With the previous equation, we obtain the numerical solutions represented in figures (\ref{fig7}) for the scalar field and for the gauge field.

%

We recall that the magnetic field and the energy can be obtained by analyzing eq. (\ref{magnetic}) and eq. (\ref{Eb1}). 

\begin{figure}[h!]
\centering
\includegraphics[scale=0.4]{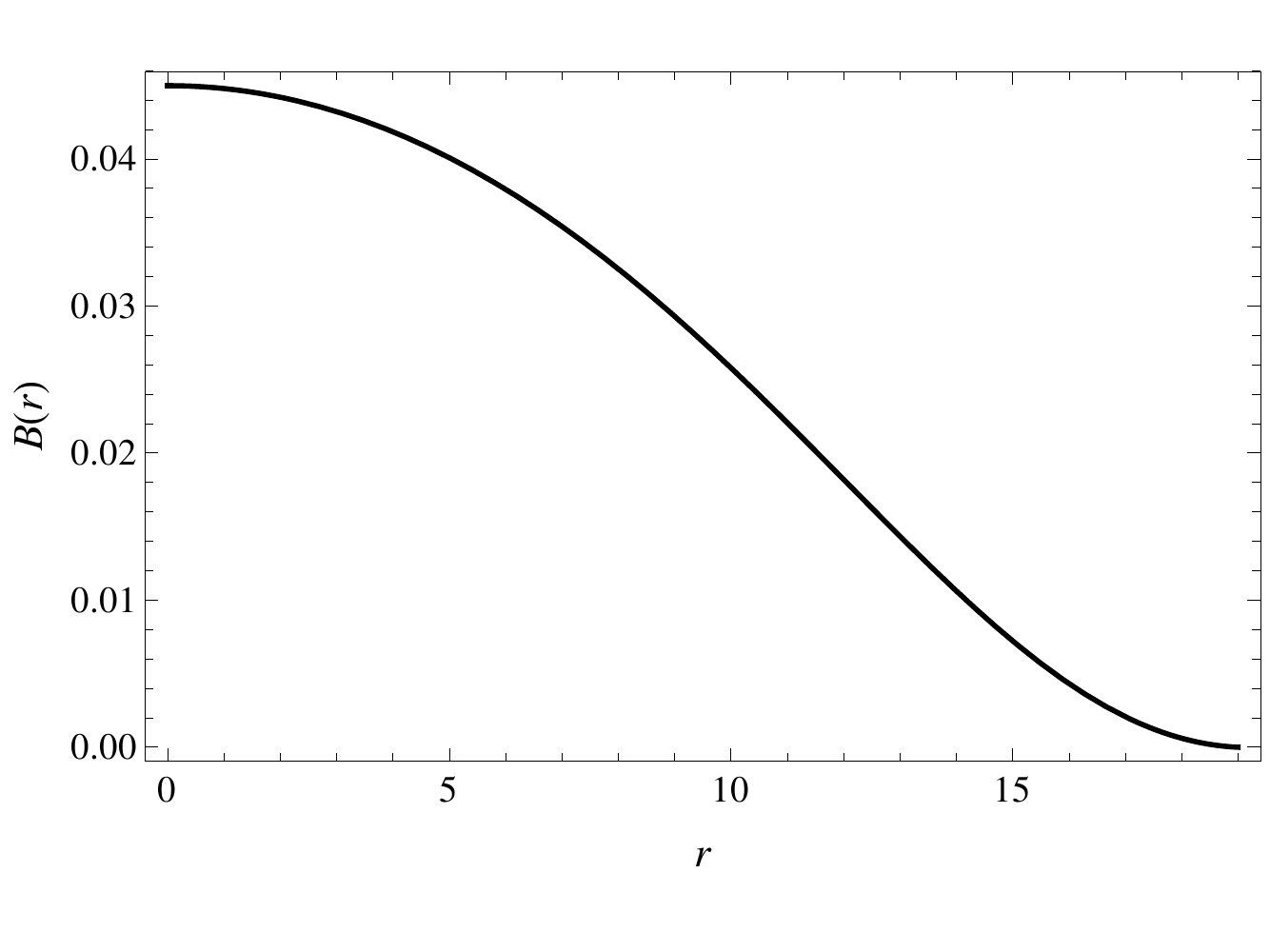}
\includegraphics[scale=0.4]{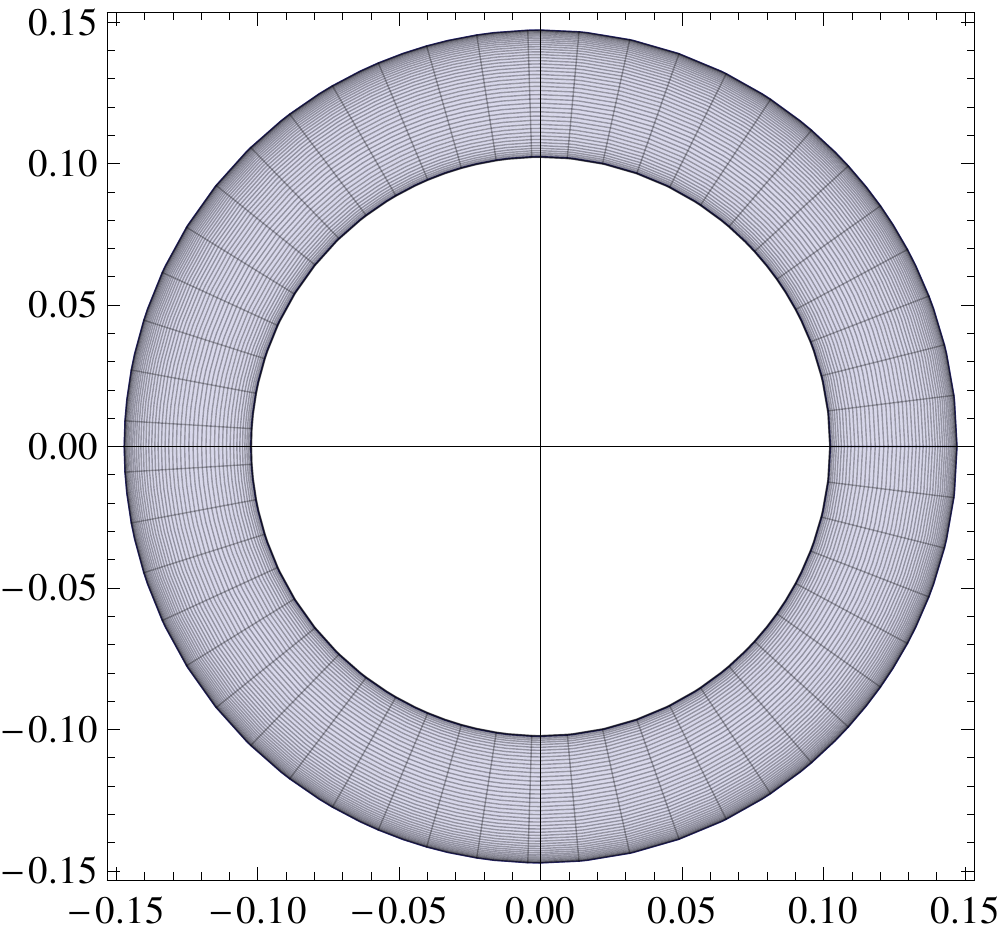}
\vspace{10pt}
\caption{The magnetic field as a function of $r$ without generalization (left side). Representation on the plane (right side).} \label{fig9}
\end{figure}
\begin{figure}[h!]
\centering
\includegraphics[scale=0.4]{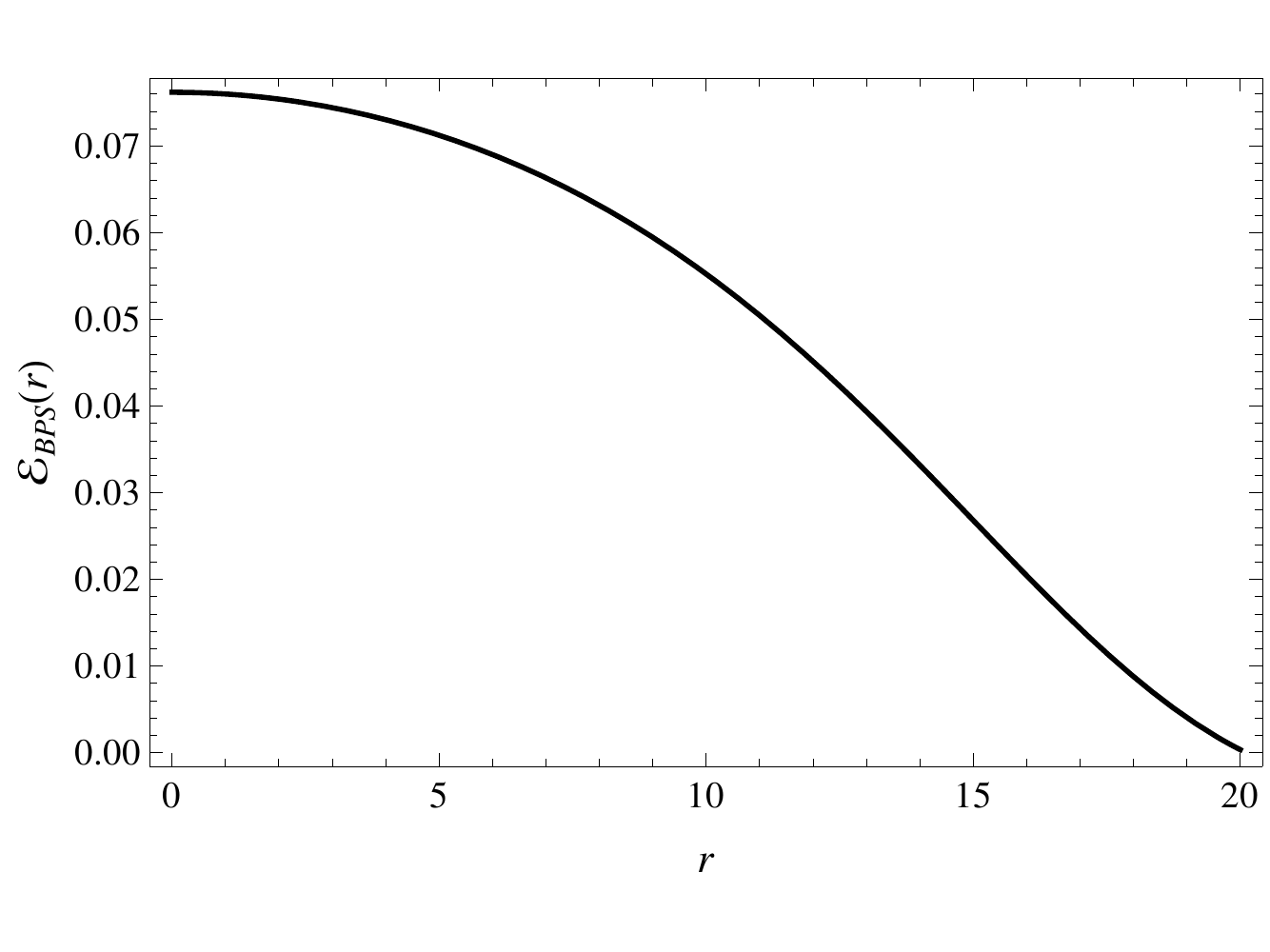}
\includegraphics[scale=0.4]{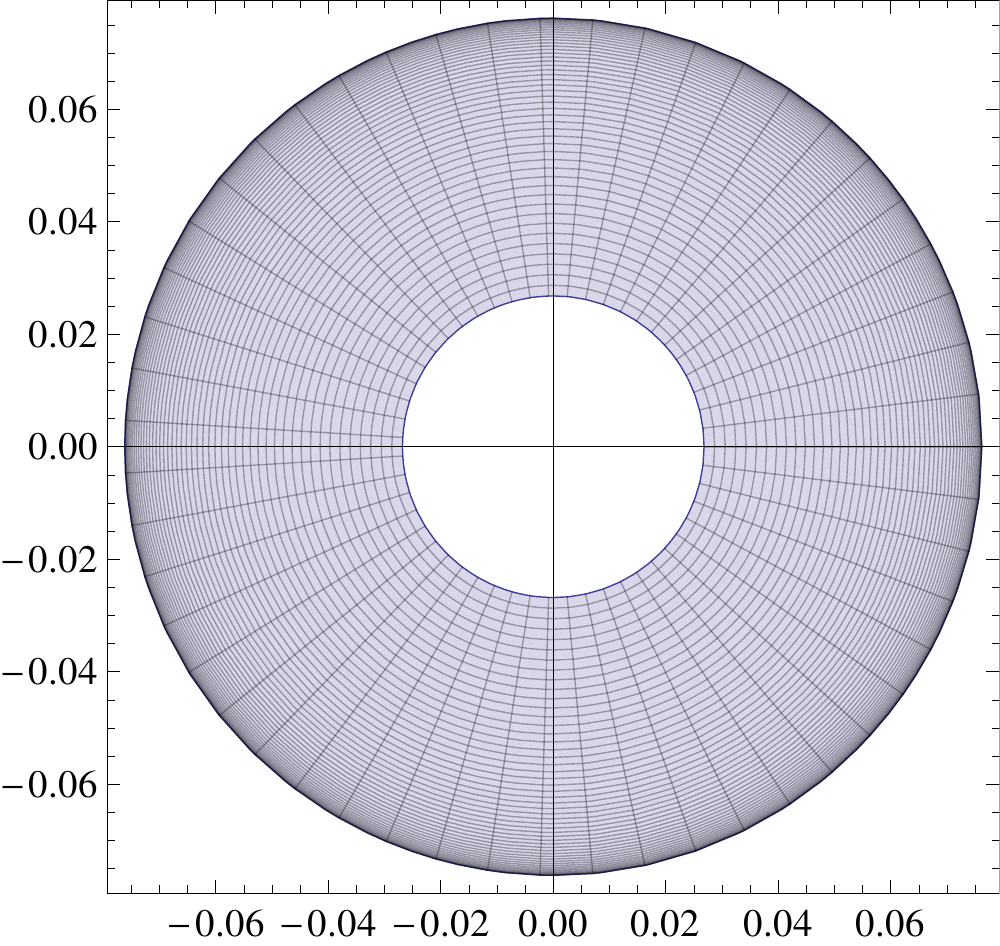}
\vspace{10pt}
\caption{The BPS energy as a function of $r$ of the vortex without generalization (left side). Representation on the plane (right side).} \label{fig10}
\end{figure}

Analyzing the Higgs field solution and the gauge field in fig. (\ref{fig7}) it was again possible to obtain the magnetic field and energy represented by figs. (\ref{fig9}) and (\ref{fig10}). We observed the existence of a hole around $r=0$. We noticed that in the case that the function $G=1$, the magnetic field and the energy present a less compacted shape. Therefore, we observe that the field has a lower intensity  near of the origin, once we are in the absence of the contribution of the logarithmic coupling. 

\section{Concluding remarks}
In this work, we study a Maxwell theory generalized by a logarithmic scalar function. In the model a gauge field is  governed by a Maxwell term, but there is an coupling between the gauge field and a scalar one. In this model, the existence of vortices solutions was well observed, with a gauge field generating a very intense magnetic field in the neighborhood of $r=0$ and with quantized magnetic flux. The fact that the intensity of the magnetic field and the energy density are very intense is due to the contribution of the logarithmic term being dominant in the model. We also obtain that the Bogomol'nyi energy of the model is quantized and given by $E_{BPS}=2\pi|n|\nu^{2}$, that is, the solutions of this model are degenerated $\nu^{2 }$ in a given sector. Through numerical analysis we obtain planar solutions, investigating the density of BPS energy and the magnetic field of the model associated with topological solutions. They were characterized as ringlike vortices. Finally,  when analyzing the particular case, i. e., when  the function $G=1$, we notice that the Higgs field and the gauge field presents a more compact shape. Therefore, for an unmodified Maxwell term, the vortex will have less energy and consequently, a less intense magnetic field.
\section{Acknowledgments}

The authors thank the Conselho Nacional de Desenvolvimento Cient\'{\i}fico e Tecnol\'{o}gico (CNPq), grant n$\textsuperscript{\underline{\scriptsize o}}$ 308638/2015-8 (CASA), and Coordena\c{c}ao de Aperfei\c{c}oamento do Pessoal de N\'{\i}vel Superior (CAPES) for financial support. FCEL would like to thank D. F. S. Veras for fruitful discussions.


\end{document}